\documentclass[fleqn,usenatbib]{mnras}

\usepackage{newtxtext,newtxmath}

\usepackage[T1]{fontenc}
\usepackage{subcaption}

\DeclareRobustCommand{\VAN}[3]{#2}
\let\VANthebibliography\thebibliography
\def\thebibliography{\DeclareRobustCommand{\VAN}[3]{##3}\VANthebibliography}

\usepackage{graphicx}	
\usepackage{amsmath}	

\usepackage{booktabs}
\usepackage{tabu}
\usepackage{natbib}
\usepackage{journal_names}
\usepackage{amsfonts}

\usepackage[nolist,nohyperlinks]{acronym}
\usepackage[acronym]{glossaries}
\usepackage{ulem}
\usepackage{lipsum}

\title[Discovery of a new Intermediate Polar counterpart to an Einstein Probe source]{Optical spectroscopic and photometric classification of the X-ray transient EP240309a (EP J115415.8-501810) as an intermediate polar }

\author[Potter et al.]{
Stephen. B. Potter,$^{1,2}$\thanks{E-mail: sbp@saao.ac.za}
David A.~H. Buckley,$^{1,3,4}$
S. Scaringi,$^{5}$
I.M. Monageng,$^{1,3}$
Okwudili D. Egbo,$^{1,3}$
\newauthor
Philip A. Charles,$^{4,6}$
N. Erasmus,$^{1}$
Carel van Gend, $^{1}$
Egan Loubser,$^{1}$
Keegan Titus,$^{1}$
Kathryn Rosie,$^{1}$
\newauthor
Hitesh Gajjar,$^{1}$
H. L. Worters,$^{1}$
Sunil Chandra,$^{1,7}$
R. P. M. Julie,$^{8}$
and Moloko Hlakola$^{1}$
\\
$^{1}$South African Astronomical Observatory, PO Box 9, Observatory, 7935, Cape Town, South Africa\\
$^{2}$Department of Physics, University of Johannesburg, PO Box 524, Auckland Park 2006, South Africa\\
$^{3}$Department of Astronomy, University of Cape Town, Private Bag X3, Rondebosch 7701, South Africa\\
$^{4}$Department of Physics, University of the Free State, PO Box 339, Bloemfontein 9300, South Africa\\
$^{5}$Centre for Extragalactic Astronomy, Department of Physics, Durham University, South Road, Durham, DH1 3LE\\
$^{6}$Department of Physics and Astronomy, University of Southampton, Southampton SO17 1BJ, UK\\
$^{7}$Center for Space Research, North-West University, Potchefstroom 2520, South Africa \\
$^{8}$South African Radio Astronomy Observatory, 2 Fir Street, Observatory 7925, South Africa \\
}

\date{Accepted XXX. Received YYY; in original form ZZZ}

\pubyear{2024}

\begin{document}
\label{firstpage}
\pagerange{\pageref{firstpage}--\pageref{lastpage}}
\maketitle

\newacronym{utc}{UTC}{Coordinated Universal Time}
\newacronym{saao}{SAAO}{South African Astronomical Observatory}
\newacronym{shoc}{SHOC}{Sutherland High-speed Optical Camera}
\newacronym{bjd}{BJD}{Barycentric Julian Date}
\newacronym{tdb}{TDB}{Barycentric Dynamical Time}
\newacronym{jd}{JD}{Julian Date}
\newacronym{gaia}{GAIA}{Global Astrometric Interferometer for Astrophysics}
\newacronym{gps}{GPS}{Global Positioning System}
\newacronym{poets}{POETS}{Portable Occultation, Eclipse, and Transit Systems}
\newacronym{moris}{MORIS}{Massachusetts Institute of Technology Optical Rapid Imaging System}
\newacronym{mit}{MIT}{Massachusetts Institute of Technology}
\newacronym{idl}{IDL}{Interactive Data Language}
\newacronym{salt}{SALT}{Southern African Large Telescope}
\newacronym{hrs}{HRS}{High-Resolution Spectrograph}
\newacronym{rss}{RSS}{Robert Stobie Spectrograph}
\newacronym{het}{HET}{Hobby Eberly Telescope}
\newacronym{bvit}{BVIT}{Berkeley Visible Image Tube}
\newacronym{salticam}{SALTICAM}{SALT Imaging Camera}
\newacronym{pipt}{PIPT}{Principal Investigator Proposal Tool}
\newacronym{rass}{RASS}{ROSAT All Sky Survey}
\newacronym{vla}{VLA}{Very Large Array}

\begin{abstract}

We report on optical follow-up observations of an X-ray source initially detected by the Einstein Probe mission. Our investigations categorize the source as an intermediate polar, a class of magnetic cataclysmic variables, exhibiting an orbital period of 3.7614(4) hours and a white dwarf spin period of 3.97 minutes. The orbital period was identified through TESS observations, while our high-speed photometric data, obtained using the 1.9m and Lesedi 1.0m telescopes at the South African Astronomical Observatory, revealed both the spin and beat periods. Additionally, we present orbitally phase-resolved spectroscopic observations using the 1.9m telescope, specifically centered on the H$\beta$ emission line, which reveal two emission components that exhibit Doppler variations throughout the orbital cycle.

\end{abstract}

\begin{keywords}
binaries: general -- cataclysmic variables -- binaries: close -- stars: individual: EP J115415.8-501810
\end{keywords}


\section{Introduction}

\cite{Lingetal2024} reported the detection of a new, highly variable X-ray source, designated EP240309a/EP J115415.8-501810, by the Einstein Probe (EP) mission \citep{2022hxga.book...86Y}, using the Wide-field X-ray Telescope (WXT) on board, during a calibration observation on March 9, 2024. The source exhibited significant variability, with its $0.5-4$ keV flux ranging between $5 \times 10^{-12}$ to $7 \times 10 ^{-12}$ erg cm$^{-2}$ s$^{-1}$ , until March 16, 2024. It was subsequently detected by the Follow-up X-ray Telescope (FXT) on board EP, on March 16, 2024. Historical observations were examined, revealing a potential identification with XMMSL J115415.6-501758, a source previously detected by XMM-Newton and Swift/XRT, albeit with variable flux levels across observations. Consistent with the FXT observations, eROSITA detected a faint source, 1eRASS J115415.7-501758, in its first six months of the all-sky survey (Merloni et al. 2024). \cite{Lingetal2024} also noted a highly variable, bright UV counterpart (Gaia DR3 5370642890382757888) suggesting that it is likely of Galactic origin and potentially a candidate Cataclysmic Variable (CV) thereby encouraging followup observations.


On March 30, 2024 \cite{Chnagetal2024} conducted follow-up radio observations at 1.28 GHz, but no radio counterpart was found within the error circle of the Einstein Probe X-ray position, nor at the position of the bright UV source.

\cite{Rodriguez2024Atel} analyzed the Gaia XP spectrum and the ASAS-SN light curve, reporting erratic variability in the observed data. Based on these observations, they suggested that the object (hereafter refereed to as EP240309a) in question is a magnetic Cataclysmic Variable.

\cite{BuckleyetalEP240309aAtel2024} conducted observations of EP240309a using the Southern African Large Telescope (SALT; \cite{Buckley2006}) on March 19, 2024. Their findings revealed a spectrum characterized by broad emission lines, echoing the observations made with the lower resolution spectrum from Gaia \citep{Rodriguez2024Atel}. The spectrum displays a steeply rising blue continuum, prominently featuring strong Balmer and HeI lines, alongside HeII 4686 and the Bowen fluorescence CIII/NIII blend (4640-4650A). {\cite{BuckleyetalEP240309aAtel2024} also reported the detection of a 3.762 h period in the TESS light curve, consistent with an orbital period.

\section{observations}
\label{sect:observations}

$V$ and $g$ filtered observations were obtained from the ASAS-SN Sky Patrol Photometry Database, see \citealp{Hartetal2023} and \citealp{Shappeeetal2014}. The observations cover a time-span of $\sim$8.3 years with a cadence of $\sim $1 measurement every few days (Fig.~\ref{figure: ASASSN  }).

Optical photometric observations of EP240309a were
obtained using the Transiting Exoplanet Survey Satellite (TESS) that cover 25 days (2019 Mar 26 - 2019 Apr 22) with a cadence of $\sim $50 data points per day (Fig.~\ref{figure: TESS }). There also exists an available ELEANOR light curve \citep{eleanorTESS2019} which was downloaded from the Mikulski Archive for Space Telescopes
(MAST).

Following the transient alert \citep{Lingetal2024} observations were promptly secured through the SAAO Intelligent Observatory programme \citep{potter:icalepcs2023-mo1bco01}. We obtained high cadence (10 s) i$^\prime$ and R filtered photometric observations with the 1.9-m and 1.0-m Lesedi telescopes of the South African Astronomical Observatory using the SHOCNWONDER and Mookodi instruments (see Table~\ref{table:observations}) in March and April 2024. Differential photometry was derived (Fig.~\ref{figure:2Apr_shoc}) using the Tea-Phot data reduction package \citep{TeaPhot2019}.

As originally reported in \cite{BuckleyetalEP240309aAtel2024}, SALT spectroscopy was obtained using the Robert Stobie Spectrograph (RSS; \citealt{2003SPIE.4841.1463B,2003SPIE.4841.1634K}) in long-slit low-resolution mode (R$\sim$ 800), with an exposure time of 1200 s, covering a wavelength range of 3500-7500\AA (see Table~\ref{table:observations}).  Data reduction was carried out using the PYSALT\footnote{For more details on pysalt visit \url{http://pysalt.salt.ac.za/}.}  software package 
 \citep{2010SPIE.7737E..25C}. The spectrum is displayed in Fig.~\ref{figure:SALTSpec} where the two gaps are due to the
detector consisting of a mosaic of three chips which in turn
results in two small gaps in the wavelength dispersion direction. Relative flux calibration was achieved using sensitivity functions derived from observations of the spectrophotometric standard star HILT 600 (HD 289002).

Spectroscopic observations were obtained using the SPUPNIC instrument \citep{spupnic2019} of the SAAO 1.9m telescope. A total of $\sim$15.8 hours spread over 4 nights with a 500s exposure time. Grating 4 was used with a dispersion of 1.3 \AA/pixel covering a wavelength range of 4200–5000\AA, \  utilizing a slit width of 1.5arc-seconds and centered on the H$\beta$ emission line (see Table~\ref{table:observations}). Comparison arc (Cu–Ar) spectra were taken at regular intervals for wavelength calibration, however the data has not been flatfielded or flux calibrated. Data reductions proceeded using standard IRAF\footnote{IRAF is distributed by the National Optical Astronomy Observatories,  which  are operated by  the Association of Universities for Research in Astronomy, Inc., under cooperative agreement with the NSF.} routines.


\begin{table*}
\caption{Log of observations.}
\label{table:observations}
\begin{center}
\begin{tabular}{l c c c c c c c c} \hline \hline
Date    & $BJD-2400000.0(TDB)$   & Telescope  & Instrument  & Filter/ & Exposure & Total time   \\
    & start   &   &  & Grating Res. &  & hr    \\ \hline \hline
2024/03/19 & 60388.55226(JD) &SALT & RSS & $R\sim 800$ & 1200s & 1200s   \\
2024/03/24 & 60394.29718435 &1.9m SAAO & SPUPNIC & 1.3\AA/pix & 21x500s & 3.25   \\
2024/03/26 & 60396.2379825 &Lesedi (1.0m) SAAO & Mookodi & i' & 10s & 2.45  \\
2024/03/28 & 60398.28792416 &1.9m SAAO & SPUPNIC & 1.3\AA/pix & 26x500s & 3.62  \\
2024/03/29 & 60399.30361589 &1.9m SAAO & SPUPNIC & 1.3\AA/pix & 20x500s & 2.71  \\
2024/04/01 & 60402.26767327 &1.9m SAAO & SPUPNIC & 1.3\AA/pix & 45x500s & 6.30  \\
2024/04/02 & 60403.33377779 &1.9m SAAO & SHOCNWONDER & R & 10s & 6.74  \\
\hline \hline \hfill
\end{tabular}
\end{center}
\end{table*}




\section{results}
\label{sect:results}

\subsection{Photometry: ASAS-SN}

Fig. \ref{figure: ASASSN } shows the archival $V$ and $g$ filtered observations from the ASAS-SN survey spanning $\sim$8.3 years. Originally reported in \cite{Rodriguez2024Atel} as showing erratic variability between V=17.2 mag and 14.4 mag, upon closer examination the light-curve appears to have a bi-modal distribution of variability. This can typically be understood as characteristic behavior of magnetic Cataclysmic Variables during transitions in accretion states, although it is more frequently observed in the synchronized polar subclass and is also prevalent in nova-like variables. The origin of high and low luminosity/accretion states is not yet fully understood and it has been suggested that such changes are due to star-spots and solar-type magnetic cycles in the donor star (e.g. \citealt{LivioPringle1994, Kafkaetal2005})

\subsection{Photometry: TESS}

Despite the low cadence, the "banding" visible in the light curve Fig.~\ref{figure: TESS } is indicative of a periodic signal. This is confirmed with a Lomb-Scargle Periodogram, Fig.~\ref{figure: TESS }, showing a dominant frequency at 6.3806(6) c/d (3.7614(4) h) and its harmonic. Errors in the recovered period were determined via bootstrapping the TESS light curve with replacement. We identify this as the orbital period of EP240309a, in agreement with the original report in \cite{BuckleyetalEP240309aAtel2024}.

\subsection{Photometry: Lesedi \& 1.9m}

Both i$^\prime$ and $R$ light curves (e.g. Fig.~\ref{figure:2Apr_shoc}) exhibit variability on the scale of hours, as well as more frequent fluctuations, consistent with flickering as seen in most CVs \citep{2014MNRAS.438.1233S}. To analyze these variations, we conducted a Lomb-Scargle Periodogram analysis on the aggregated light curves (see Fig.~\ref{figure:LombScar}).

At the lower frequency end, we detected peaks corresponding to the orbital period and its second and third harmonics. Additionally, the periodogram revealed a frequency of 362.5 c/d (equivalent to 3.972 min), which we associate with the spin period of the white dwarf, as denoted by the blue dashed line in Fig.~\ref{figure:LombScar} and its top inset.  The top inset of Fig.~\ref{figure:LombScar} 
also features a green dashed line indicating the beat frequency  ($\omega-\Omega$)  which coincides with a prominent peak in the periodogram. Moreover, the periodogram depicted in the lower right inset reveals the presence of ($2\omega$) and ($2\omega-2\Omega$) frequencies.

We also observe a notable frequency around 60.5 c/d, roughly in-between 9$\Omega$ and 10$\Omega$, which does not correspond to any harmonics of the identified frequencies. Further observations are necessary to determine whether this represents a transient signal.

In summary, we have identified the orbital and spin frequencies as 6.3806(6) c/d and 362.5(1) c/d, respectively, based on data from TESS (for the orbit) and combined observations from Lesedi and the 1.9m telescopes (for the spin). The uncertainties in the spin frequency, indicated in parentheses, was measured from the full width at half maximum (FWHM) of the periodogram peak, taking into account potential aliases.

In Fig.~\ref{figure:2Apr2024_trail}, we present the spin and beat-folded light curves derived from the more extensive 1.9m photometric observations, illustrated in the lower left and right panels, respectively. Both types of light curves reveal the presence of single pulses. The upper panels of Fig.~\ref{figure:2Apr2024_trail} depict the evolution of consecutive spin and beat-folded light curves throughout the orbital cycle. It is important to note that each successive folded light curve has undergone normalization to mitigate the dominant higher-amplitude orbital modulation. Within the constraints of the signal-to-noise ratio of our observations, we observe that the spin pulse remains relatively stable in phase throughout the orbital cycle. In contrast, the beat pulse evolves to progressively earlier beat phases as the orbital cycle progresses. This observation suggests that the emission source responsible for the high-frequency pulses is anchored in the frame of the rotating white dwarf, rather than resulting from reprocessed emission originating from a location locked in the orbital frame. The latter scenario would result in a beat pulse that remains constant over the orbital cycle.

A detailed examination of orbital phases $\sim$0.1-0.2, as shown in the top panels of Fig.~\ref{figure:2Apr2024_trail}, uncovers what seems to be a swift advancement of both the spin and beat pulses by a complete cycle. This observation may suggest that, for the majority of the time, a single spin-dominating pulse is present, likely originating near an accreting magnetic pole of the white dwarf. However, it appears that, on occasion, accretion may transition to the opposite magnetic pole, causing a pulse that exhibits a shift by $\sim$half a spin/beat cycle and then another $\sim$half a spin/beat cycle when reverting back to the original pulse.


\begin{figure}
    \includegraphics[width = 0.5\textwidth]{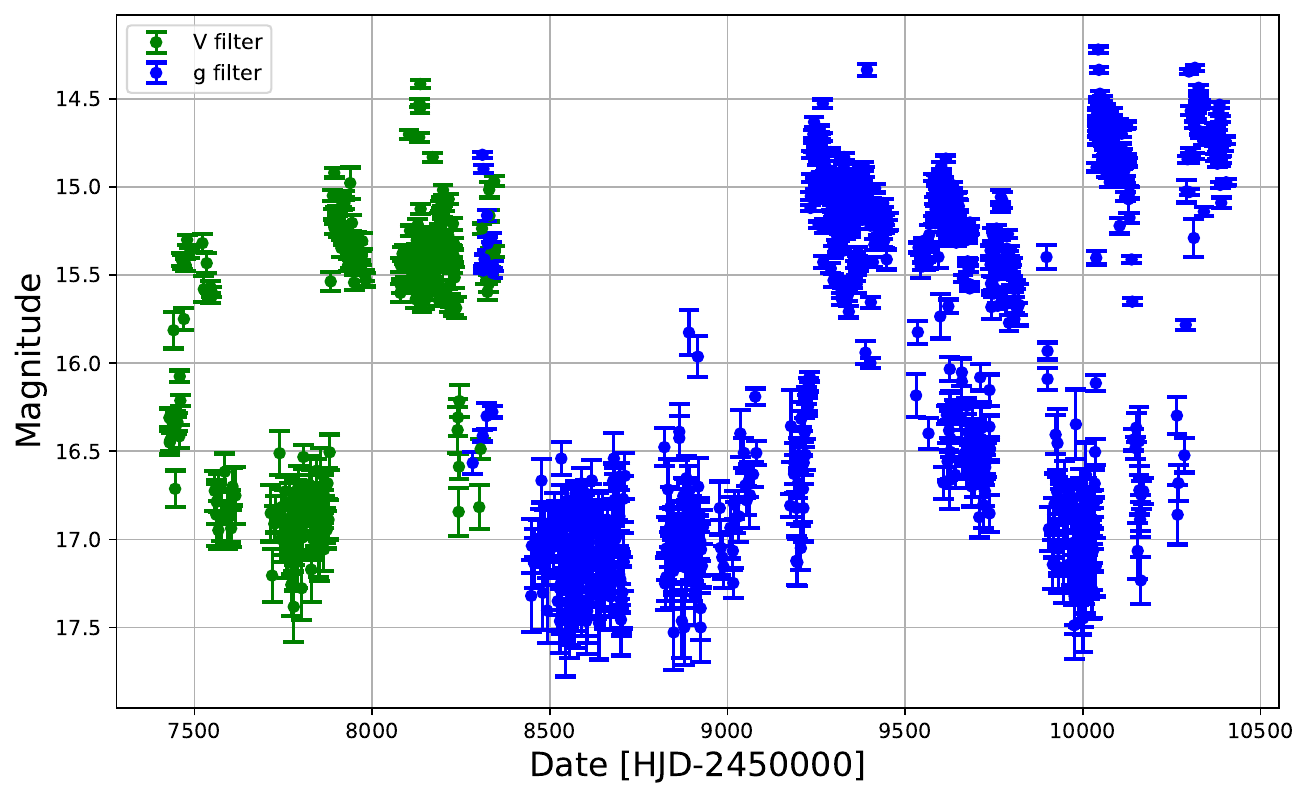}
    \caption{V and g filtered light curves from ASAS-SN.}
    \label{figure: ASASSN }
\end{figure}

\begin{figure}
    \includegraphics[width = 0.5\textwidth]{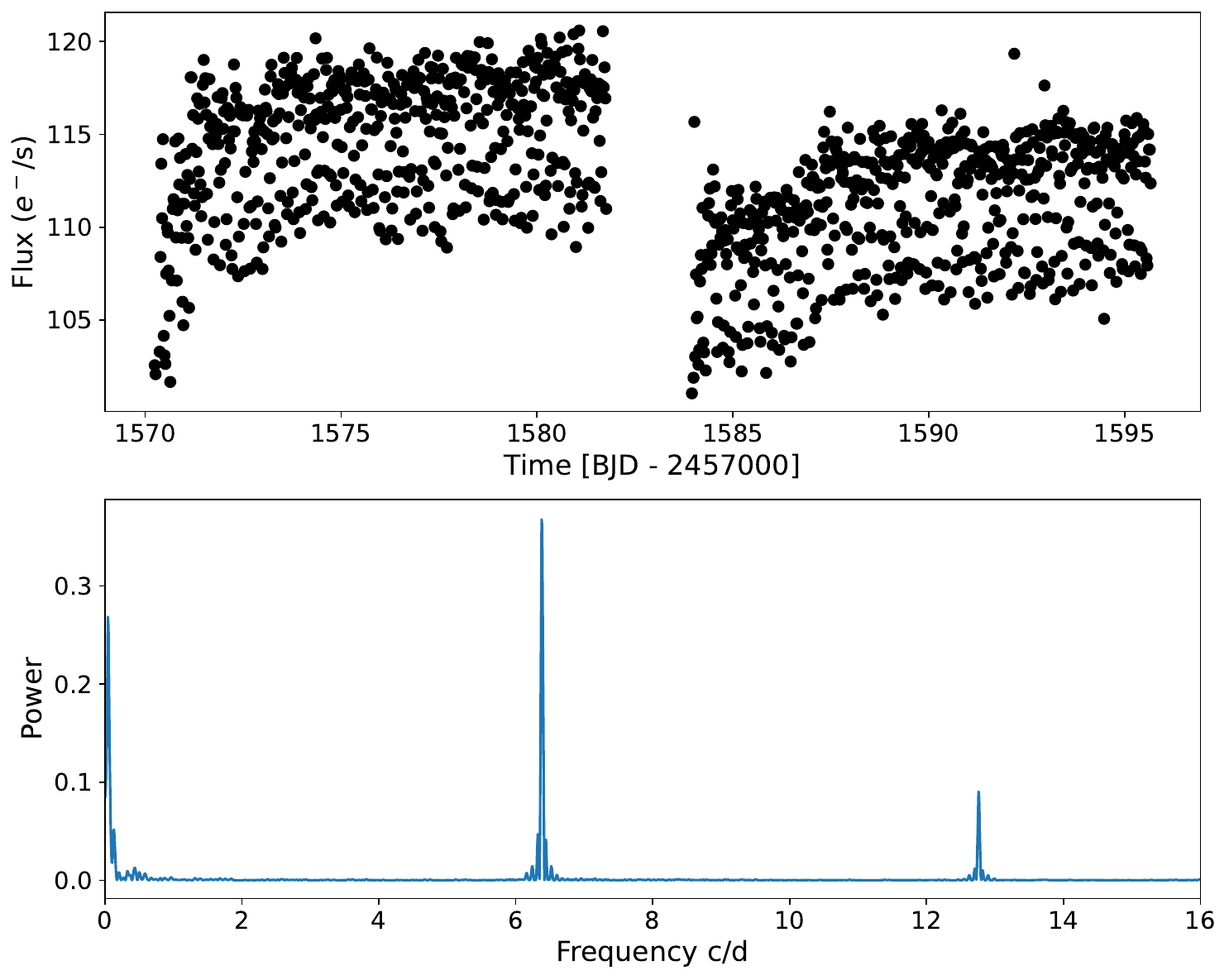}
    \caption{Upper plot: The TESS light curve of EP240309a spanning $\sim$ 25 days. Lower plot: The corresponding Lomb-Scargle Periodogram of the mean subtracted curve.}
    \label{figure: TESS }
\end{figure}


\begin{figure}
    \includegraphics[width = 0.5\textwidth]{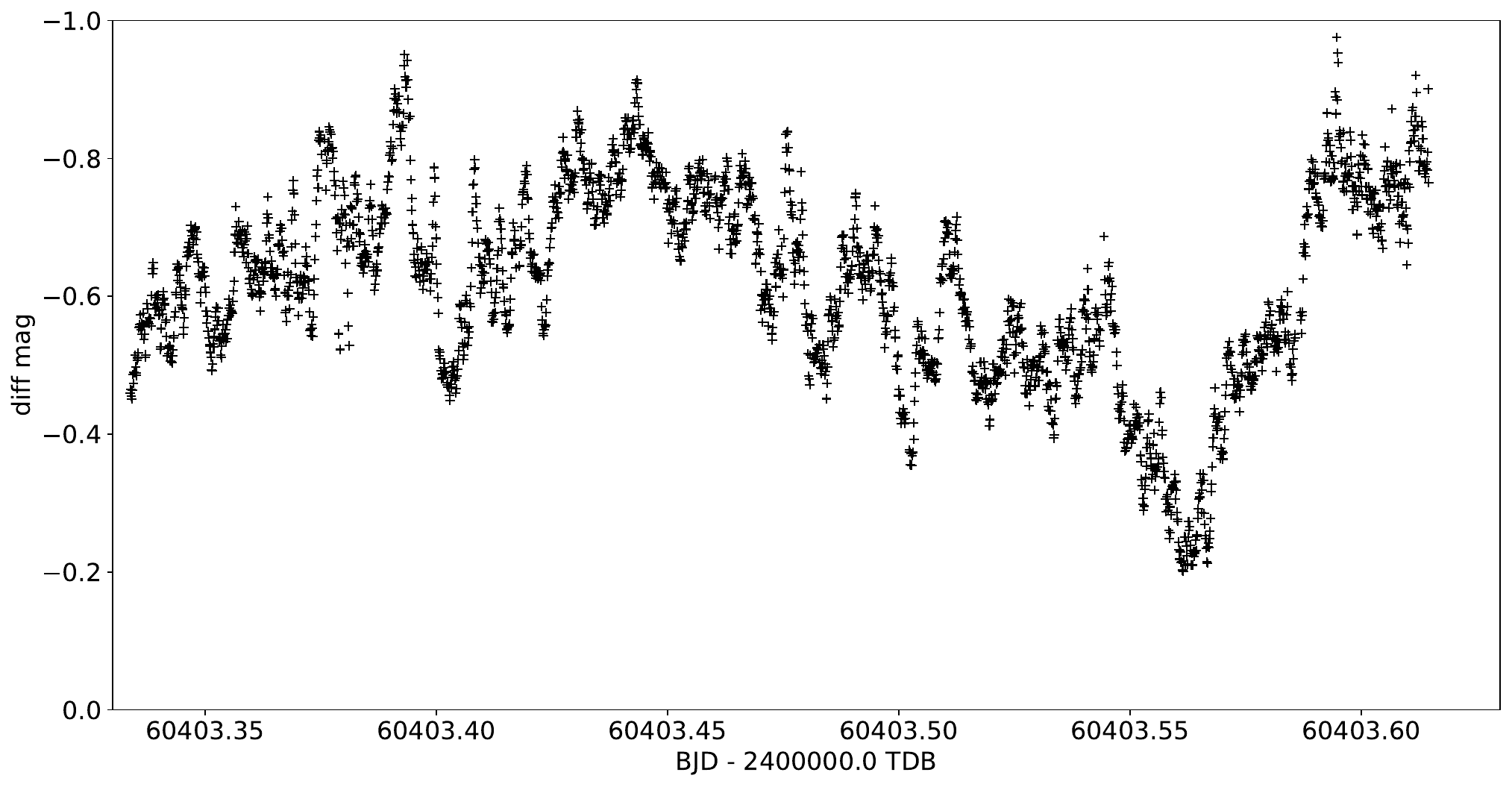}
    \caption{Example photometry, showing R filtered photometric observations made with the shocnwonder instrument on the SAAO 1.9m telescope spanning 6.74 hours on 2 April 2024.}
    \label{figure:2Apr_shoc}
\end{figure}

\begin{figure}
    \includegraphics[width = 0.5\textwidth]{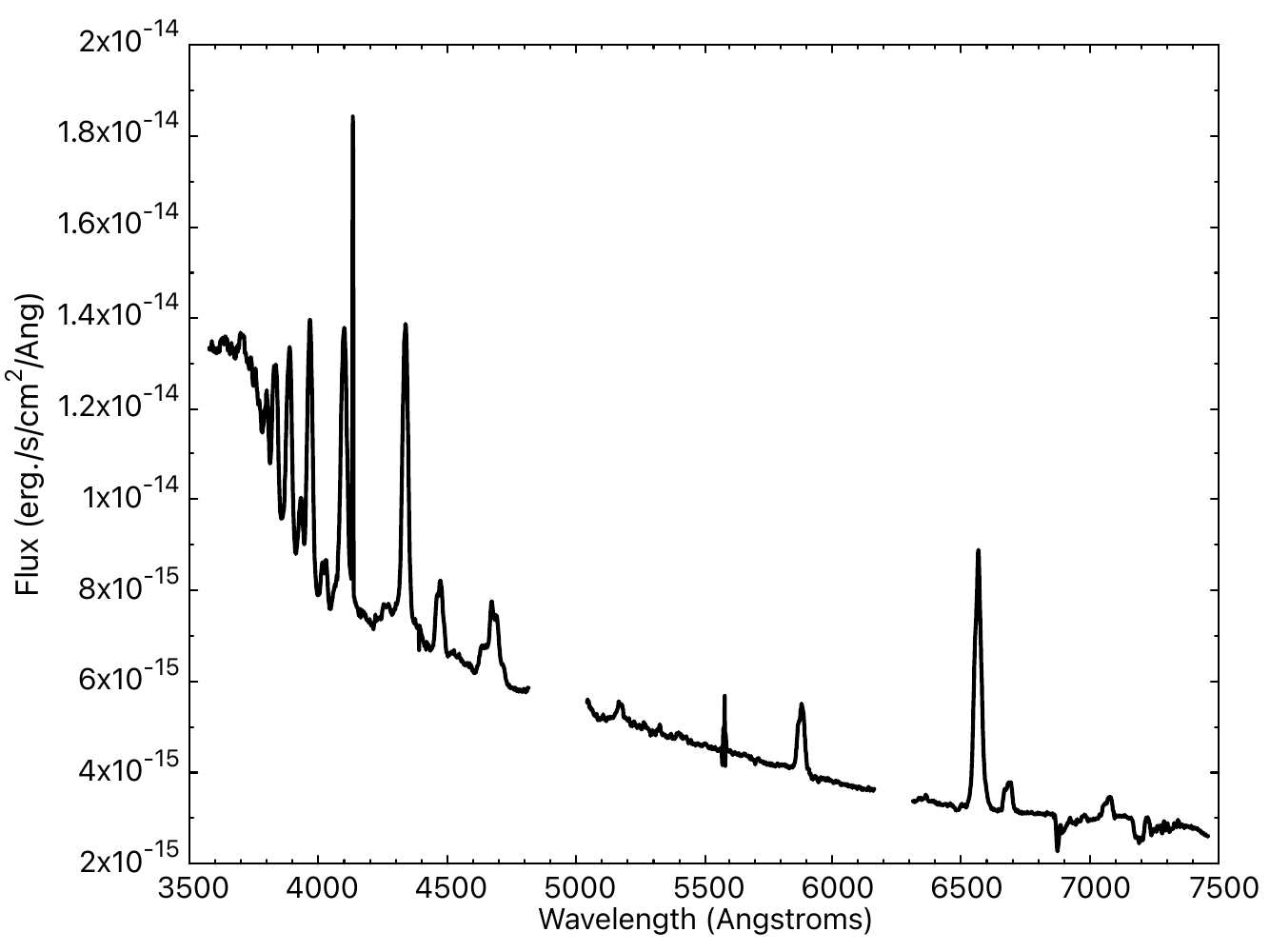}
    \caption{Optical spectrum of EP240309a from SALT}
    \label{figure:SALTSpec}
\end{figure}



\begin{figure*}
    \includegraphics[width = \textwidth]{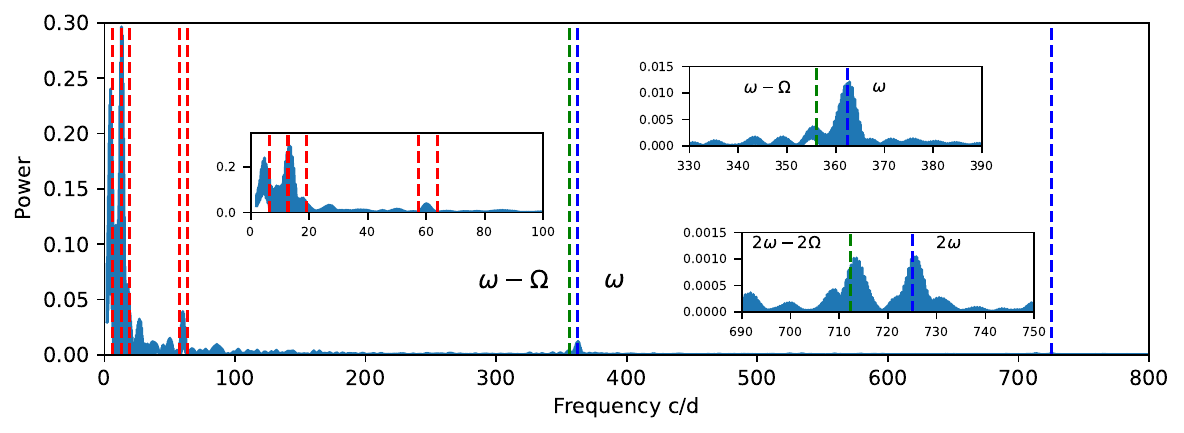}
    \caption{Lomb-Scargle Periodogram of the combined 3.25hr and 6.74hr i$^{\prime}$,R filtered photometry. Periodgrams of individual light curves show similar results (not shown). Assumed $\Omega = 6.3796 = 3.762$hr. Left inset: Red dashed lines correspond to $\Omega, 2\Omega, 3\Omega, 9\Omega, 10\Omega$. Top inset: Blue dashed line indicates possible $\omega = 362.5 = 3.97$mins and green indicates $\omega-\Omega$. Note the peak between frequencies $9\Omega$ and $10\Omega$ also present on individual nights (not shown.)}
    \label{figure:LombScar}
\end{figure*}

\begin{figure}
    \includegraphics[width = 0.5\textwidth]{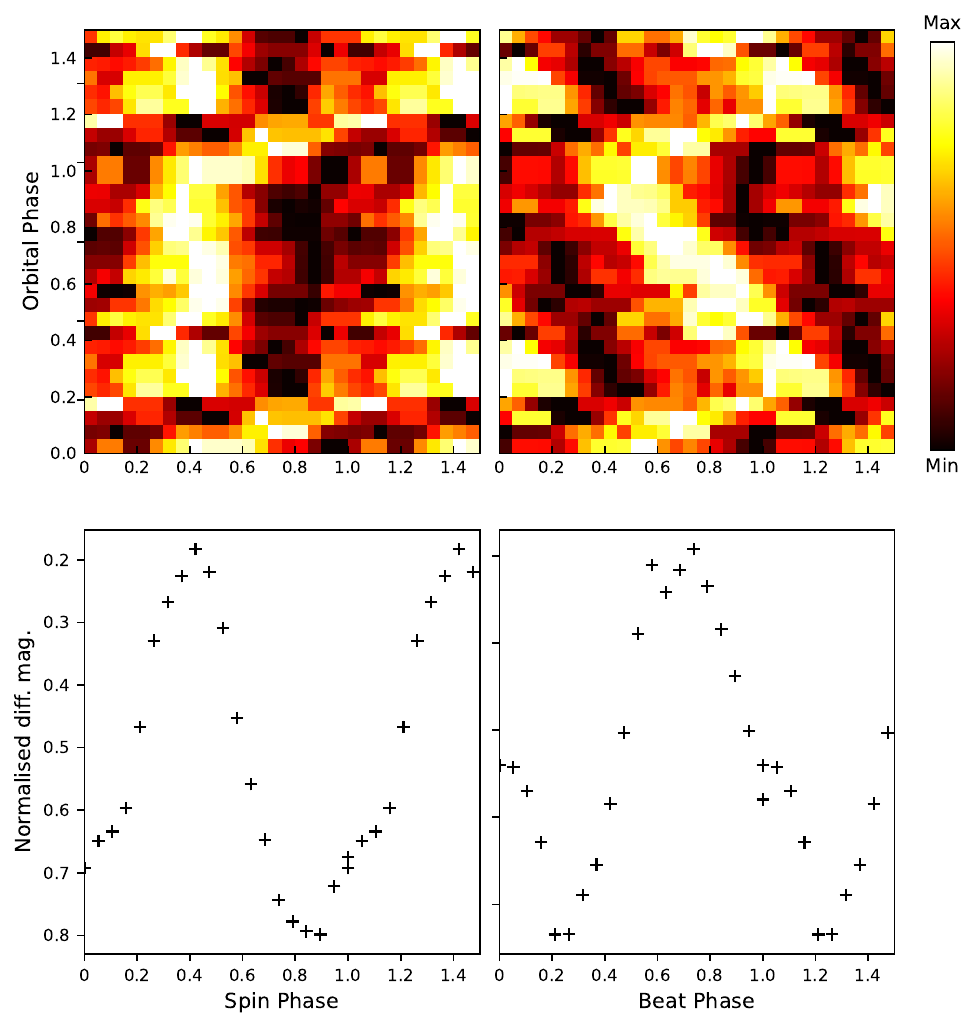}
    \caption{Top left: Evolution of the spin pulse over the orbital period. Top right: Evolution of the beat pulse over the orbital period. Bottom left: 1D normalised spin-folded and binned light curve. Bottom right: 1D normalised beat-folded and binned light curve.}
    \label{figure:2Apr2024_trail}
\end{figure}

\begin{figure}
    \includegraphics[width = 0.5\textwidth]{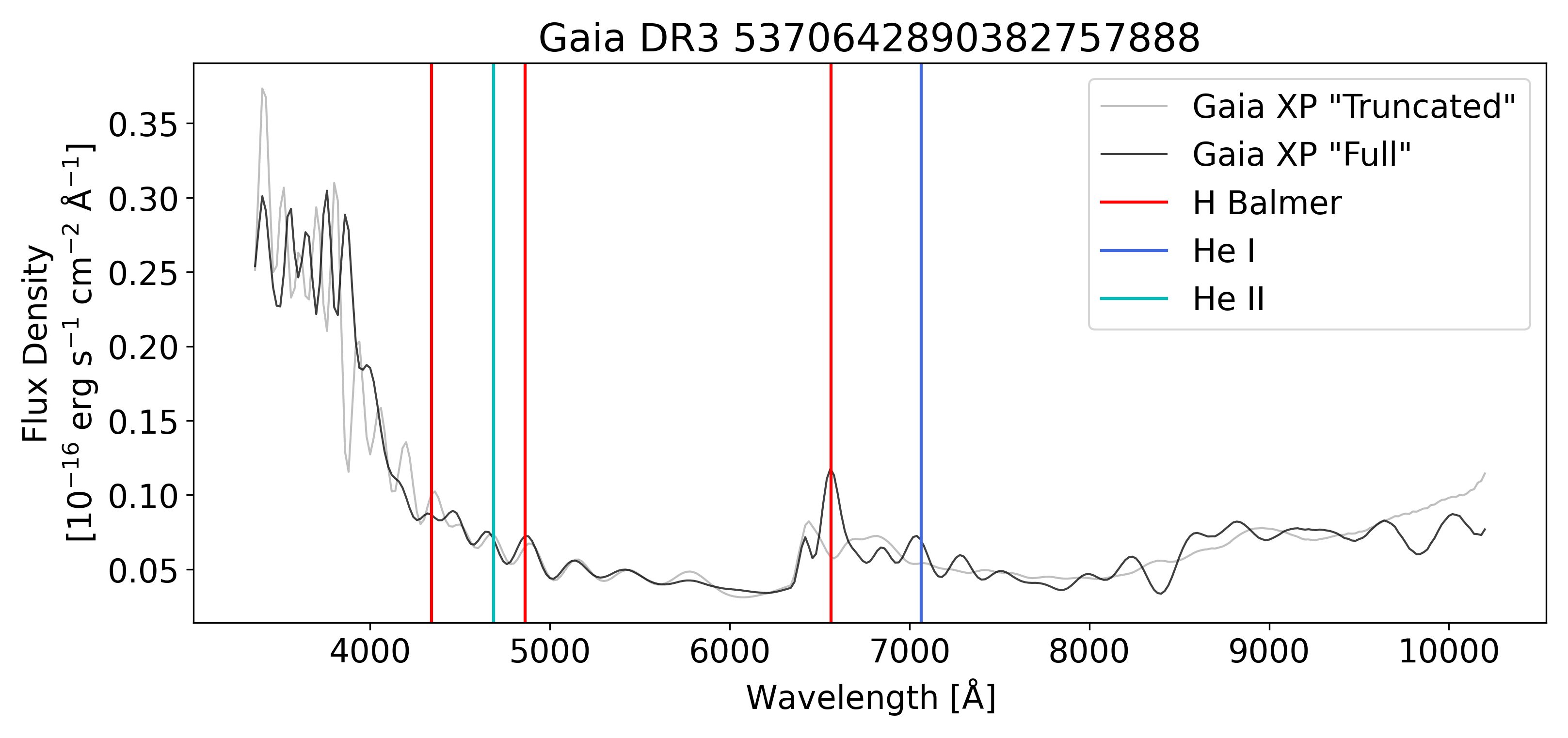}
    \caption{Gaia XP Continuous spectrum of EP240309a, showing strong emission lines on a blue continuum, typical of a CV.}
    \label{figure:Gaia-XP-spec}
\end{figure}

\subsection{Gaia: astrometry and spectrum}
As reported in \cite{Rodriguez2024Atel}, EP240309 is a Gaia \citep{2016A&A...595A...1G} source in the Data Release 3 catalogue 
(DR3 5370642890382757888; \cite{2023A&A...674A...1G}), at a magnitude of $G$ = 16.2. Astrometry for the source gives a parallax of 3.2041 mas
and proper motion of 68.270 mas/yr. The distance determined from \cite{2021AJ....161..147B} is $d$ = 309.5 $\pm$ 0.4 pc.

Furthermore, Gaia produced a low resolution (R = 30$-$100) XP Continuous spectrum, which is shown in Fig, ~\ref{figure:Gaia-XP-spec}. The spectrum is typical of a cataclysmic variable, suggested by \cite{Rodriguez2024Atel} (notwithstanding the low spectral resolution)  to possibly be a magnetic CV, based on the presence of HeII 4686\AA\ , a high excitation line which is prominent in these relatively X-ray luminous CV systems.

\subsection{Spectroscopy: SALT}

Pronounced broad emission lines are seen, similar to what is reported for the lower resolution Gaia spectrum \citep{Rodriguez2024Atel}, on a steeply rising blue continuum (see Fig,~\ref{figure:SALTSpec}). In addition to strong Balmer and HeI lines (4471, 5018, 5876, 6678, 7065\AA), the HeII 4686 and Bowen fluorescence CIII/NIII blend (4640-4650\AA) are also prominent. Due to the chip gap in the RSS detector, the H$\beta$ line was missed. H$\alpha$ is characterized by a FWHM of around 1500 km/s.

\subsection{Spectroscopy: SAAO, 1.9m}

The left panel of Fig.~\ref{fig:comparison_plots} displays the phase-folded spectroscopic observations from the SAAO 1.9m telescope, centered on the H$\beta$ emission line. Despite the low signal-to-noise ratio in individual spectra, it is evident from this figure that H$\beta$ is composed of multiple components exhibiting Doppler variations throughout the orbital cycle. Upon careful examination, we tentatively discern two distinct components, which are highlighted by dashed curves in the right panel of Fig.~\ref{fig:comparison_plots} to guide the eye. Although the H$\beta$ Doppler tomogram, which is not depicted here, reveals two areas of intensified emission, the absence of an accurately determined orbital ephemeris precludes the correct phasing and therefore the identification of these features in relation to the binary system's components. The full width at half maximum (FWHM) of H$\beta$ is approximately 1500 km/s, corroborating findings from an earlier SALT observation. 

\begin{figure}
    \centering
    \begin{subfigure}[b]{0.5\columnwidth}
        \includegraphics[width=\linewidth]{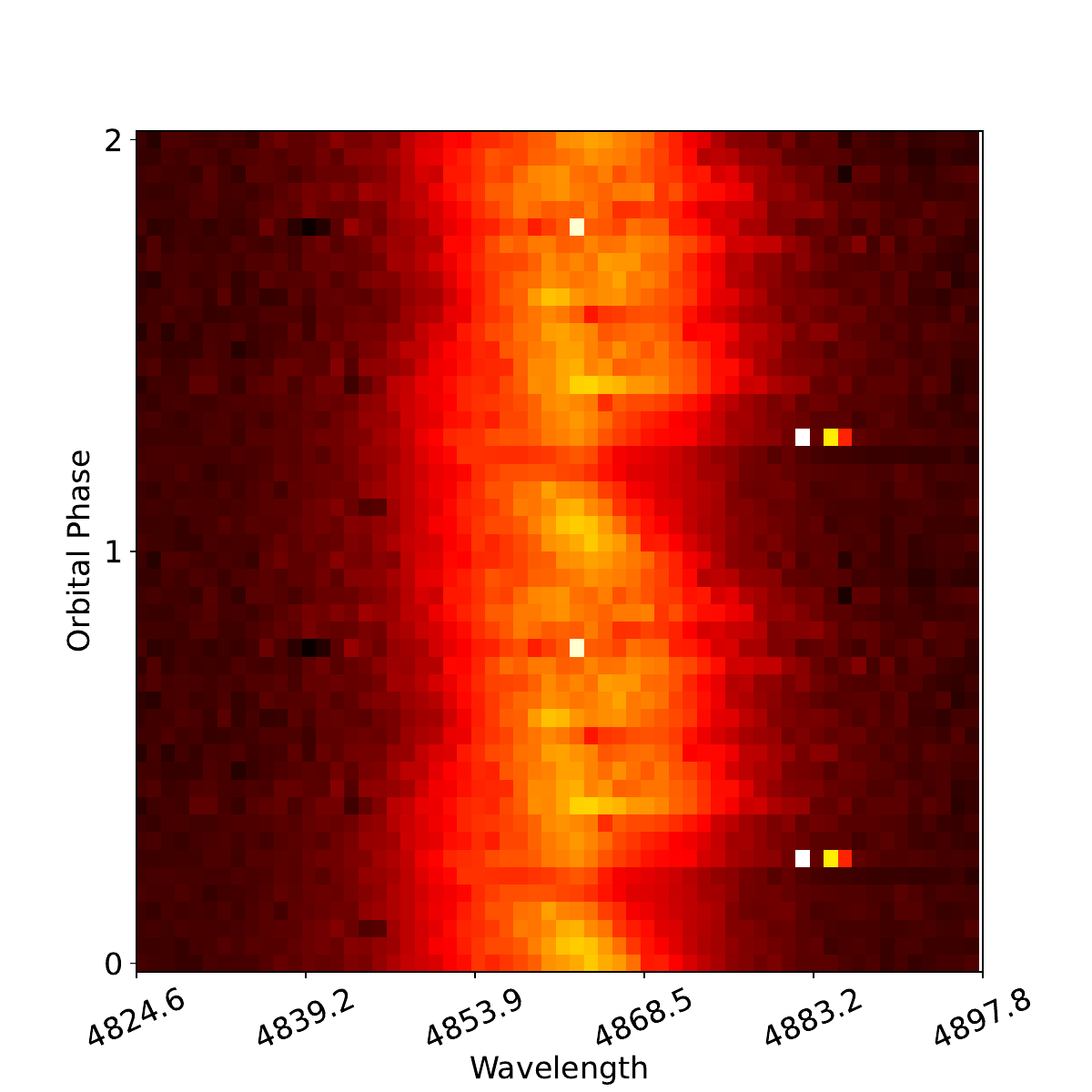}
    \end{subfigure}%
    \hfill
    \begin{subfigure}[b]{0.5\columnwidth}
        \includegraphics[width=\linewidth]{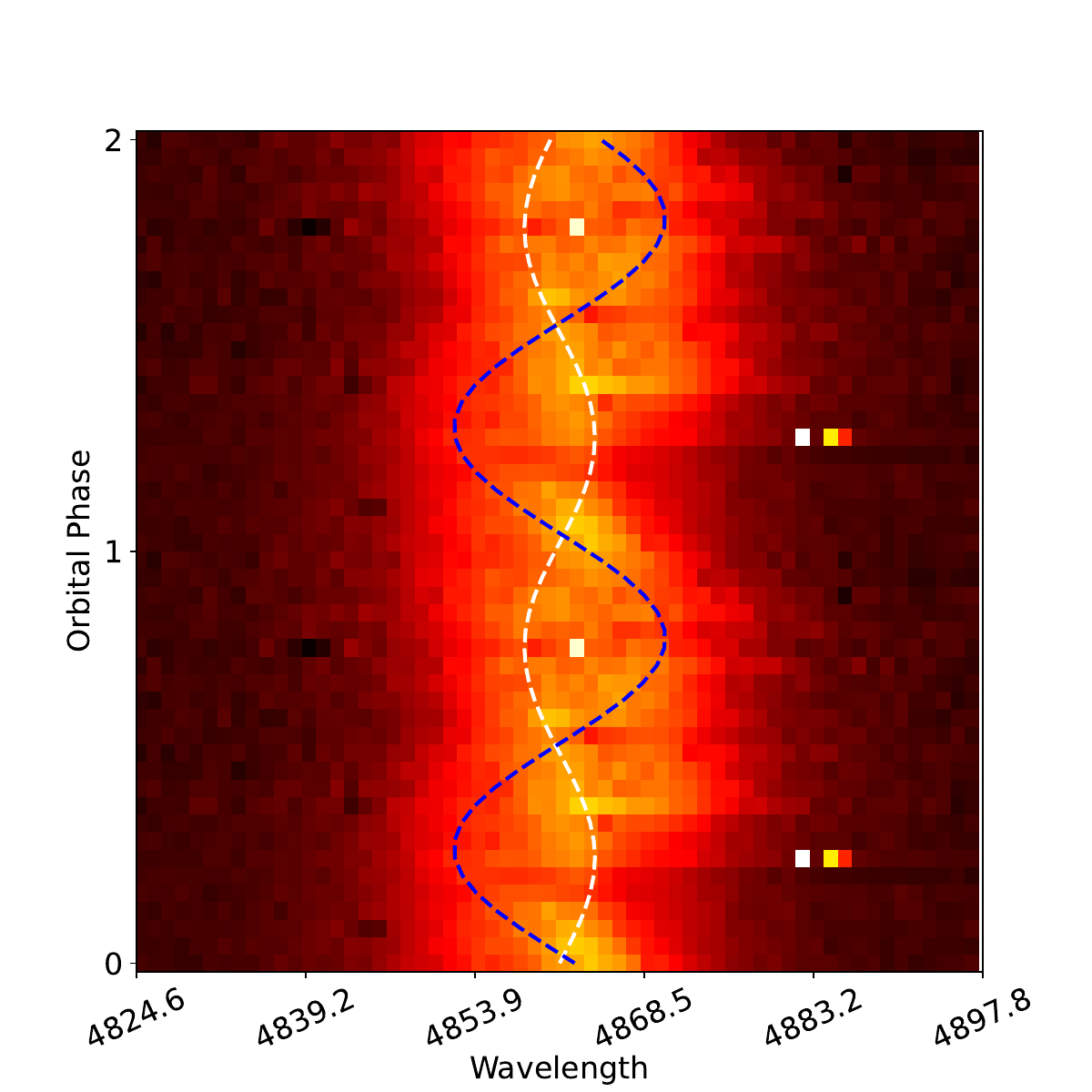}
    \end{subfigure}
    \caption{Left plot shows all the spupnic observations of EP240309a centered on H$\beta$ folded and binned on the assumed orbital period of 3.764hr. Right plot: same as left but overlaid with two sine functions indicating two possible components.}
    \label{fig:comparison_plots}
\end{figure}



\section{Summary and Discussion}
\label{sect:summary}

\begin{figure}
    \includegraphics[width = 0.5\textwidth]{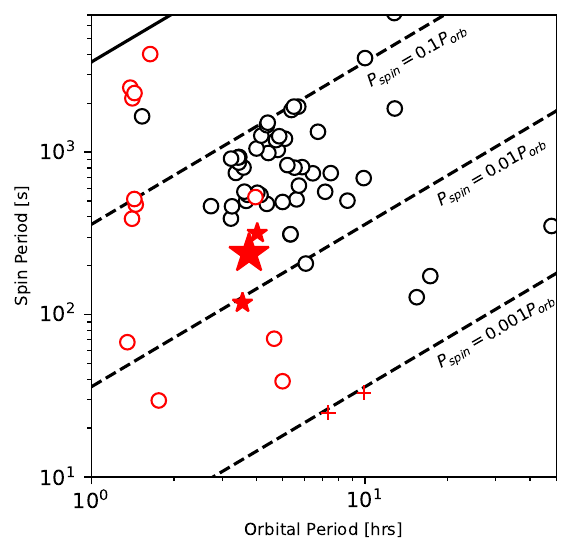}
    \caption{Orbital versus spin period of the so-called ironclad intermediate polars, based on data compiled by Koji Mukai (\url{https://asd.gsfc.nasa.gov/Koji.Mukai/iphome/iphome.html}) The large red star represents EP240309a, small red stars denote the two known white dwarf pulsar systems, red crosses denote the two propeller systems (AE Aqr and LAMOST J024048.51+195226.9) and open red circles illustrate Low Luminosity IPs  \citep{LLIP}. }
    \label{figure:PspinPorb}
\end{figure}

Prompted by the detection of a transient X-ray source, EP240309a, through the Einstein Probe mission, we embarked on an optical follow-up campaign. A Lomb-Scargle period analysis of TESS observations disclosed a predominant period of 3.7614(4) hours, indicating the likely binary orbital period. This was corroborated by our high-speed photometric observations at SAAO, which additionally unveiled frequencies consistent with a white dwarf spin period of 3.97 minutes and the corresponding beat frequency. These findings categorize the source as an intermediate polar.

Our analysis of the long-term light curve from the ASAS-SN archive indicated almost bimodal transitions between brightness levels, a behavior typically observed during the high and low state transitions in magnetic Cataclysmic Variables (see e.g. \cite{Kafkaetal2005}) further corroborating our IP classification.

Spectroscopic examination using SALT revealed the characteristic broad emission line spectrum typical of magnetic cataclysmic variables. Moreover, time-resolved SAAO (1.9m) spectroscopic observations centered on the H$\beta$  emission identified two Doppler variations across the orbital cycle. The typical double-peaked lines emission expected in a disc accreting system (see e.g. \cite{Mhlahlo2007}) was not seen although this would be best confirmed with better signal-to-noise observations. To accurately phase our observations and determine the origin of these Doppler varying components, further time-resolved spectroscopic observations, for instance of the CaII triplet from the irradiated face of the secondary star, are required (see e.g. \cite{KhangalePotter2020}).

We encourage further more detailed followup X-ray observations in order to ascertain the X-ray spectral properties and also the spin and/or beat temporal behavior. This will help to reveal the accretion dynamics e.g. disc-fed or disc-stream overflow and also whether the system is a hard or soft X-ray source (e.g. \cite{JoshiRawat2023}, \cite{Vermetteetal2023}). Similarly time-resolved photo-polarimetry will further enhance our understanding of the accretion dynamics and the overall energetics (e.g. \cite{PotterRomero2012}).

In Fig.~\ref{figure:PspinPorb} we show the position of EP23409a in the $P_{spin} - P_{orb}$ diagram \citep{Norton2008}, together with other IPs and related objects. The four fastest spinning systems are represented by the bottom two red circles and crosses, with  CTCV J2056–3014 and V1460 Her, the left and right circles, respectively. The propeller systems, LAMOST J024048+195226 and AE Aqr, are the left and right red crosses, respectively.

It is interesting to note that the orbital period and the fast spin period of EP240309a fall midway between those of the two known so called white dwarf pulsar systems (small red stars in Fig.~\ref{figure:PspinPorb}), AR Sco \citep{Marsh2016} and J191213.72-441045 \citep{Pelisoli2023}, \citep{Schwope2023}. These two exceptional systems are characterised with properties that set them aside from all other known CV systems, namely strong pulsed emission on the white dwarf spin period, detectable from radio to X-rays including optical polarization (\cite{Buckley2017}, \cite{PotterBuckley2018}, \cite{Pelisoli2023}) with no signs of accretion. We do not yet fully understand either what evolutionary path AR Sco and J191213.72-441045 followed, or what underlying physical characteristics of the binary constituents set them so apart from other CVs and in particular intermediate polars. Given its similarities to these systems in terms of basic stellar constituents, periods, and binary size, EP240309a clearly warrants further detailed investigation.

\section*{Acknowledgements}
The spectroscopic observations
with the Southern African Large Telescope (SALT) were obtained under the programme  2021-2-LSP-001 (PI: David Buckley). Other ground-based observations were obtained with the facilities
of the SAAO Sutherland observing station.

This paper includes data collected by the TESS mission, which are publicly available from the Mikulski Archive for Space Telescopes (MAST). Funding for the TESS mission is provided by NASA’s Science Mission directorate. 

The Gaia XP Continuous spectrum was obtained from the Gaia DR3. This work has made use of data from the European Space Agency (ESA) mission
{\it Gaia} (\url{https://www.cosmos.esa.int/gaia}), processed by the {\it Gaia}
Data Processing and Analysis Consortium (DPAC,
\url{https://www.cosmos.esa.int/web/gaia/dpac/consortium}). Funding for the DPAC
has been provided by national institutions, in particular the institutions
participating in the {\it Gaia} Multilateral Agreement.

DAHB would like to thank Tony Rodriguez for useful discussions. DAHB and SBP acknowledge research support from the National Research Foundation. SS is supported by STFC grant ST/X001075/1. IMM is supported by the South African NRF and the UCT VC 2030 Future Leaders Programme.

We thank the referee for a very quick and positive response and suggesting Fig.~\ref{figure:PspinPorb}.

\section*{Data availability}

The reduced data underlying this article will be shared on reasonable
request to the corresponding author.




\bibliographystyle{mnras}
\bibliography{mnras_template} 







\bsp	
\label{lastpage}
\end{document}